\documentclass[12pt]{iopart}

\usepackage{float}
\usepackage{graphicx}
\usepackage{harvard}
\usepackage{csquotes}
\usepackage{hyperref}
\usepackage{subfig}

\begin{document}

\title[]{Fast optimal decomposed modification of FDK with potential decreasing of memory consuming}

\author{V Andriiashen$^{1,2}$, D Kozhevnikov$^1$}

\address{Dzelepov Laboratory of Nuclear Problems, Joint Institute for Nuclear Research, Joliot-Curie 6, 141980 Dubna, Russia}
\address{Department of General and Applied Physics, Moscow Institute of Physics and Technology, Institutskiy per. 9, 141700 Dolgoprudny, Russia}
\ead{vladandriyashen@gmail.com, dkozhevn@jinr.ru}
\vspace{10pt}
\begin{indented}
\item[]June 2017
\end{indented}

\begin{abstract}
We present a new algorithm for 3D cone-beam tomography. The algorithm is based on decomposition of the cone-beam backprojection operation and angular decimation. It has computational complexity of $O(N^{3.5})$ and allows considerable reduction of peak memory usage in comparison with conventional algorithms. Tests with real data demonstrate the acceleration, achieved by using our algorithm instead of FDK, with 20-fold speedup for a $2000 \times 2000 \times 720$ image. The algorithm is compared with other fast FDK algorithms.
\end{abstract}

%

\section{Introduction}

The reconstuction problem in cone-beam computed tomography is to recover a volume from a set of its line-integral projections at different angles. X-rays diverge as a cone from source and illuminate the object. Data corresponding to line integrals along these rays is recorded on a planar or cylindrical detector surface. We focus on the geometry in which the source moves around the object on a circular orbit. For this geometry the Feldkamp algorithm (FDK) \cite{fdk} is most often used in practice.

The FDK consists of 2 stages: filtering the projections and backprojection. Assuming that width of projections equals N and number of angle projections is $P = O(N)$, the convolution requires $O(N^3 \log N)$ operations when implemented using Fast Fourier transform. The computational complexity of backprojection is $O(N^4$) for all conventional algorithms.

Several algorithms were proposed to accelerate the FDK \cite{turbell}. The fast hierarchical Feldkamp  proposed by Xiao et al. \cite{rec3d:base} requires $O(N^3 \log N)$ operations instead of $O(N^4)$. It uses divide-and-conquer principle and is based on a hierarchical decomposition of a backprojection operation into sub-volumes and decimation of the projection angles.

Based on these concepts, we aimed to optimize memory consumption without considerable loss of speedup. The algorithm FDK-OD (optimal decomposition) presented in this article requires $O(N^{3.5})$ operations and allows to reduce required peak memory usage. Moreover, it requires small computational constant compared with the fast hierarchical FDK due to the smaller number of interpolation operations. FDK-OD provides 20-fold speedup for $2000 \times 2000 \times 720$ image that is still significant acceleration.

\section{Method description}
\subsection{General principles}
The FDK algorithm allows to reconstruct $N \times N \times N$ volume from P projections of size $N \times N$. $N$ denotes width and height of detector and P is number of angle projections. Projections are usually represented as a set of sinograms. Each sinogram corresponds to the data obtained from one row of detector on different rotation angles.  Every pixel of output volume is calculated as weighted sum of P values of sinograms. Total computational cost is $O(N^3 P)$ or $O(N^4)$ if $P = O(N)$.

\begin{figure}[H]
\begin{center}
\subfloat[]{
  \begin{minipage}[b]{0.38\textwidth}
    \includegraphics[width=\textwidth]{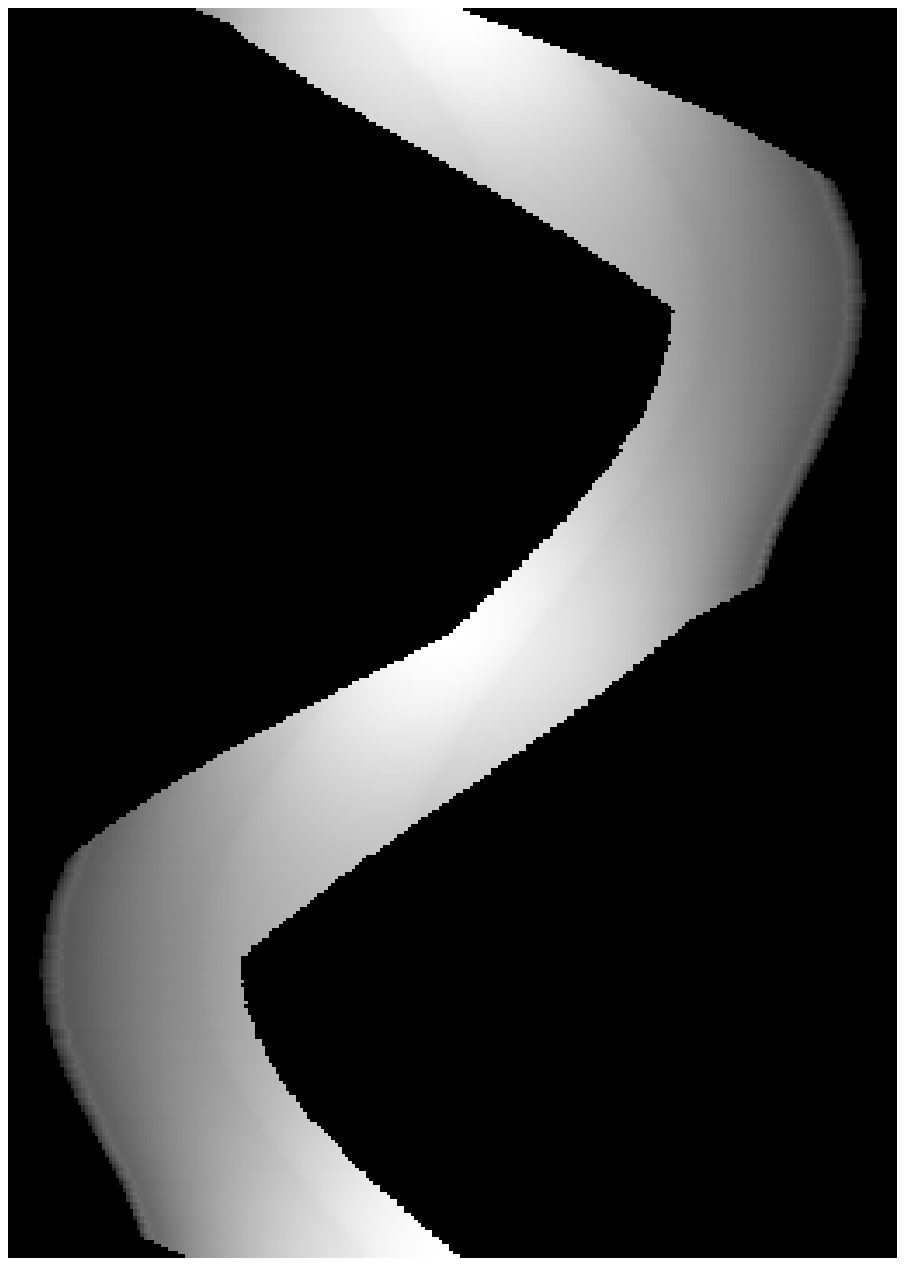}
    \label{roi_sin}
  \end{minipage}
}
\subfloat[]{
  \begin{minipage}[b]{0.535\textwidth}
    \includegraphics[width=\textwidth]{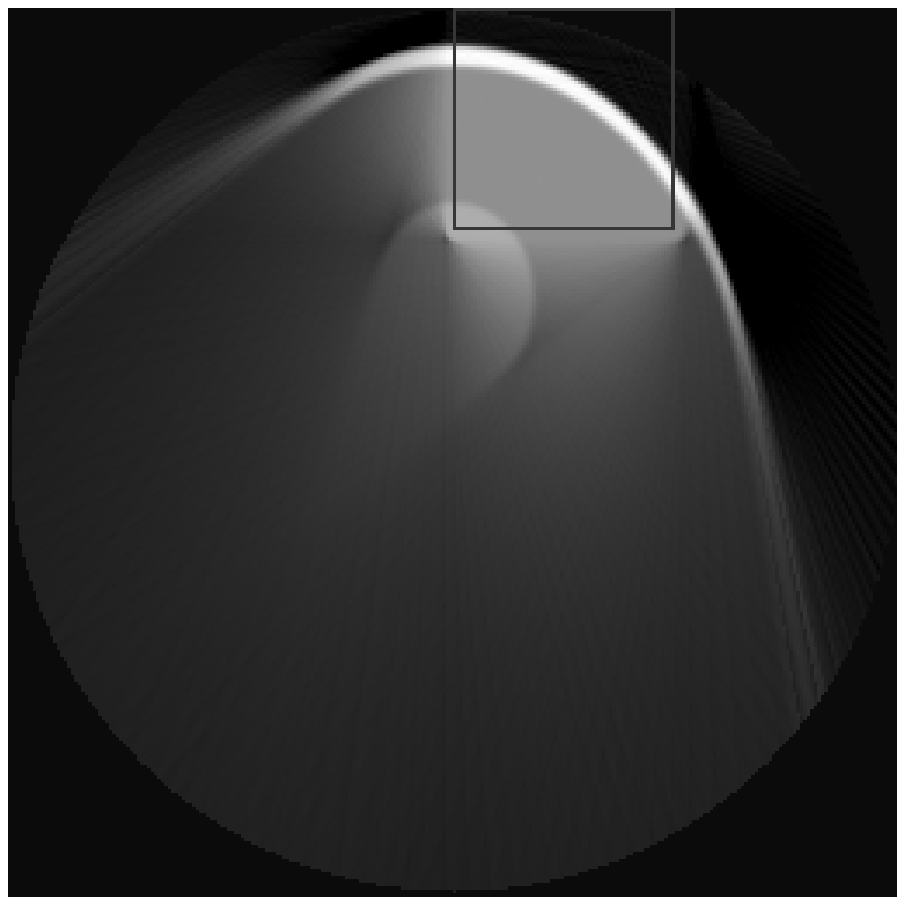}
    \label{roi_res}
  \end{minipage}
}
  \caption{Sinogram for $\frac{1}{16}$ of the image (1a) and the reconstructed region (1b)}
\end{center}
\end{figure}

Consider the process of reconstruction of a part of slice. Without loss of generality, assume that this part is a square. The FDK algorithm allows to reconstruct this square without using the whole sinogram data. For example, the part of sinogram shown in the \fref{roi_sin} allows to accurately reconstruct the marked square in the \fref{roi_res}.  Technically, this means that before the stage of reconstruction for every angle projection and sinogram number we can precompute which pixels of the row will be used in following reconstruction. For this computation one of the FDK formulas is used:

\begin{equation} \label{pix_pos}
a(x,y,\beta) = R \frac{y \cos\beta - x \sin\beta}{R + x \cos\beta + y \sin\beta}
\end{equation}

Here $a(x,y,\beta)$ denotes the position of the pixel in the row, R is a distance between the source of radiation and detector and $\beta$ is number of angle projection. (x, y) are coordinates of point of the square, so every point has corresponding value of $a(x,y,\beta)$. Minimal and maximal $a(x,y,\beta)$ give the boundaries of the part of the row that is used to reconstruct the square. Further calculations require only a minimum and maximum, so it is sufficient to compute $a(x,y,\beta)$ only for vertexes of the square.

All formulas do not include number of sinogram and boundaries remain approximately the same for all slices of sinogram data in real applications. Reconstruction computational cost is proportional to the area of the output image and the number of projections at different angles.

Consider the division of the output image f with size $N \times N$ into $4^s$ squares, where s is integer. In this decomposition each square's side is $\frac{N}{2^s}$.

The decomposition of image into squares preserves total area and doesn't lead by itself to speedup. However, acceleration can be achieved using angular decimation. The idea is to use smaller number of angle projections to reconstruct small part of the image. For parallel beam geometry it is proved by Basu et al. that angular decimation doesn't lead to significant quality loss \cite{rec2d:base,rec3d:radon}. In cone-beam geometry it is shown that this concept also can be used \cite{rec3d:base}.

Angular decimation takes as input a sequence of sinogram pixels that allow to reconstruct single output pixel. Thus length of the signal corresponds to the number of angle projections in the sinogram data. Pixel positions are calculated according to the formula (\ref{pix_pos}) and can be not integer, therefore interpolation is needed. Technically, sequences for decimation are prepared by shifting rows of sinogram, thus the centers of needed parts result to be on one column.

Angular decimation is divided in two steps: filtering and downsampling \cite{decimation}. Filtering step uses low-pass filter to prevent aliasing of the output image. This requires O(P) operations for fixed size filter. In practice, filters with small kernel can be used. Downsampling step allows to reduce the filtering cost by the factor of 2. because filtered may be calculated only in half of all points. After the decimation pixel values are returned to original positions using second shift with interpolation.

The image decomposition leads to slight quality losses. They are caused by interpolation and angular decimation. The impact of these factors can be reduced if more time consuming methods are implemented. Interpolation can be linear, cubic, spline etc. In our implementation simple linear interpolation was used. Poor angular decimation leads to aliasing artifacts in the output image. Quality can be adjusted by changing size of filter kernel and the low-pass filter implementation.

The concept of decomposition and decimation can be used to achieve $O(N^3 \log N)$ operations as in the algorithm FHBP (Fast Hierarchical Backprojection) \cite{rec3d:base}. It uses recursive decomposition. Firstly, sinogram data is decomposed to reconstruct 4 quarters of original image. If sinogram data's size is $N \times N \times P$, a computational cost is:

\begin{equation}
Dec_1 = C_d \times 4 N \frac{N}{2} P
\end{equation}

This operation produces 4 sets of decomposed data with size $N\times\frac{N}{2}\times\frac{P}{2}$. After that, decomposition into 4 parts is applied to every quarter. This leads to computational cost:

\begin{equation}
Dec_2 = C_d \times 16 N \frac{N}{4} \frac{P}{2} = Dec_1
\end{equation}

The main advantage of the FHBP is that decomposition computational cost remains the same for all stages of decomposition and this allows to reconstruct single pixel from $O(1)$ values from sinogram data:

\begin{equation}
Rec = C_r \times N^3
\end{equation}

The full reconstruction time is given by the sum of s decomposition stages and one reconstruction stage. Here s denotes the number of decomposition stages and is limited by the image size: $s \le \log_2 N$. Total amount of operations is:

\begin{equation}
Total = 2 C_d N^2 P \log_2 N + C_r N^3 \propto N^3 \log_2 N
\end{equation}

\subsection{FDK-OD - new approach}

We propose to use decomposition of original sinogram data into small parts without intermediate stages. Consider the decomposition with s stages in which image is divided into $4^s$ squares. The reconstruction computational cost:

\begin{equation} \label{fdkod-rec}
Rec = C_1 \times 4^s N \frac{N}{2^s} \frac{N}{2^s} \frac{P}{2^s} = C_1 \frac{N^4}{2^s}
\end{equation}

The decomposition computational cost:

\begin{equation} \label{fdkod-dec}
Dec = C_2 \times 4^s N \frac{N}{2^s} P = C_2 N^3 2^s
\end{equation}

The total computational cost:
\begin{equation} \label{fdkod-total}
Total = C_1 \frac{N^4}{2^s} + C_2 N^3 2^s
\end{equation}

It may be noted that number of image parts grows faster than the decomposition computational cost for every part decreases. This leads to exponential growth of operations required by the decomposition stage. Thereby, there is optimal number of stages for which sum of reconstruction and decomposition time is minimal. The minimization of (\ref{fdkod-total}) leads to $2^s = \sqrt[]{\frac{C_1}{C_2} N}$. Overall, our algorithm requires $O(N^{3.5})$ operations. 

To preserve an image quality, it is advisable to skip one or two decimation stages. This increases the total computational cost and affects the optimal number of decomposition stages, but considerably improves an image sharpness.

One important property of FDK-OD in comparison with FHBP lies in the number of interpolation operations. FHBP uses $\log_2$ N shifts interpolations, while FDK-OD requires only one interpolation. Firstly, this indicates that FDK-OD could have lower computational constant. Secondly, a big amount of interpolation operations smooths image and leads to quality losses. This can be fixed using more precise interpolation method but it leads to even greater computational constant.

A considerable advantage of FDK-OD is the possibility of significant reduction of peak memory usage. Unoptimized FDK requires all sinogram data to be stored in RAM. Full memory usage is slightly bigger and varies over the course of the algorithm workflow. Therefore, the efficient optimization should concentrate on decreasing the number of projections used at the same time. The FDK allows to precalculate the range of sinograms used to reconstruct pixel (x, y) of slice z:

\begin{equation}
b(x, y, z, \beta) = z \frac{R}{R + x\cos \beta + y\sin \beta},
\end{equation}
where $\beta$ varies from 0 to $360\deg$.

These calculations don't greatly affect the consumed time but allow to store only a fraction of sinogram data at the same time.

FDK-OD reconstructs every slice piecemeal. Reconstruction stage for every part of image requires only decimated part of sinogram that is less than original by a factor of $2^s$. Moreover, reconstruction and decomposition stages don't require the whole sinogram for their work because they need only a fraction of sinogram corresponding to reconstructed part of slice. Thus, FDK-OD is able to reconstruct image without storing the whole sinogram in memory.

This interesting property is a consequence of the original sinogram decomposition exactly to the final part without intermediate stages. FHBP on every decomposition stages stores all decimated data in memory. This is essential to obtain a computational acceleration.

\section{Performance evaluation}

Our algorithm was tested on simulated Shepp-Logan-3D head phantom and real datasets obtained from MARS scanner \cite{mars}. For performance evaluation the only difference between real and simulated data is a non-ideal geometry. This leads to extra stage of conversion from non-ideal to ideal geometry, that is general for any FDK algorithm. In our analysis only the time for reconstruction and decomposition stages are compared, because reading, an initial convolution and geometrical transformations require the same time for the FDK and FDK-OD.

\begin{figure}[!h]
\begin{center}
	\includegraphics[width=0.99\textwidth]{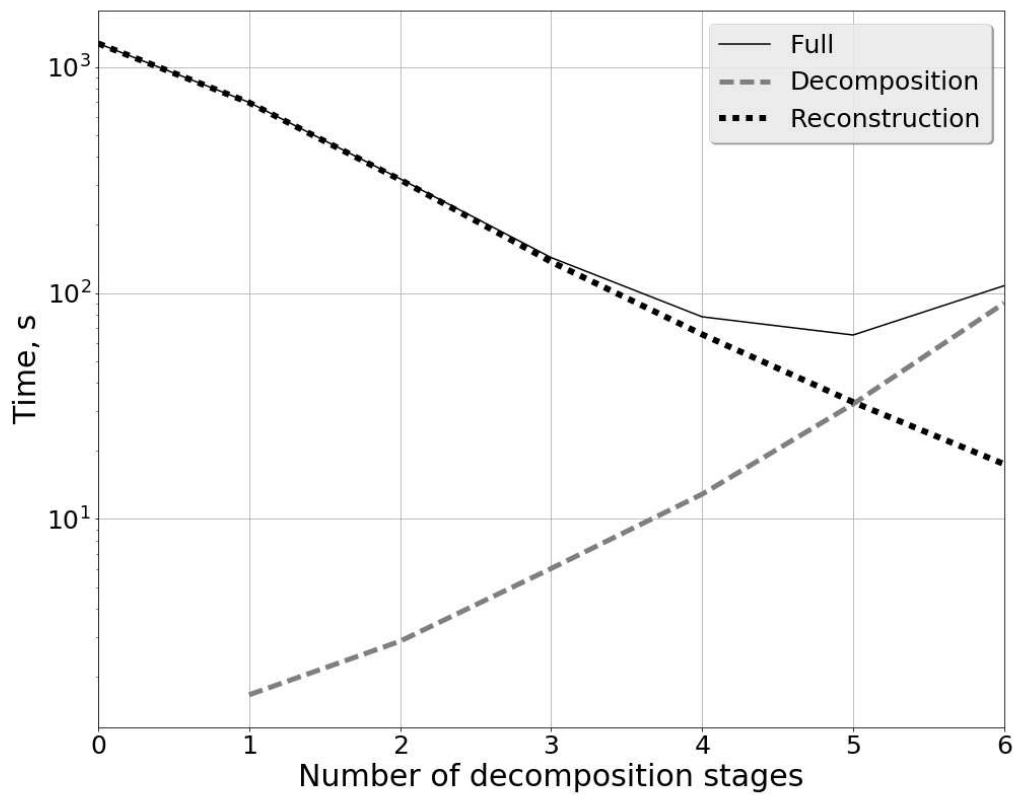}
    \caption{Time spent on decomposition and reconstruction stages}
    \label{time_decrec}
\end{center}
\end{figure}

\Fref{time_decrec} shows the dependence of time for reconstruction and decomposition stages separately and their sum on the number of decomposition stages. As predicted by theory, time spent on reconstruction decays exponentially due to the decimating procedure. Decomposition cost is negligible for small s, but growth also exponentially. This results in the existence of optimal number of stages that gives minimal time of volume reconstruction.

For performance tests we used images with N varying from 256 to 2500. Theory predicts computation cost of algorithm to be $O(N^{3.5})$.

\begin{figure}[!h]
\begin{center}
	\includegraphics[width=0.99\textwidth]{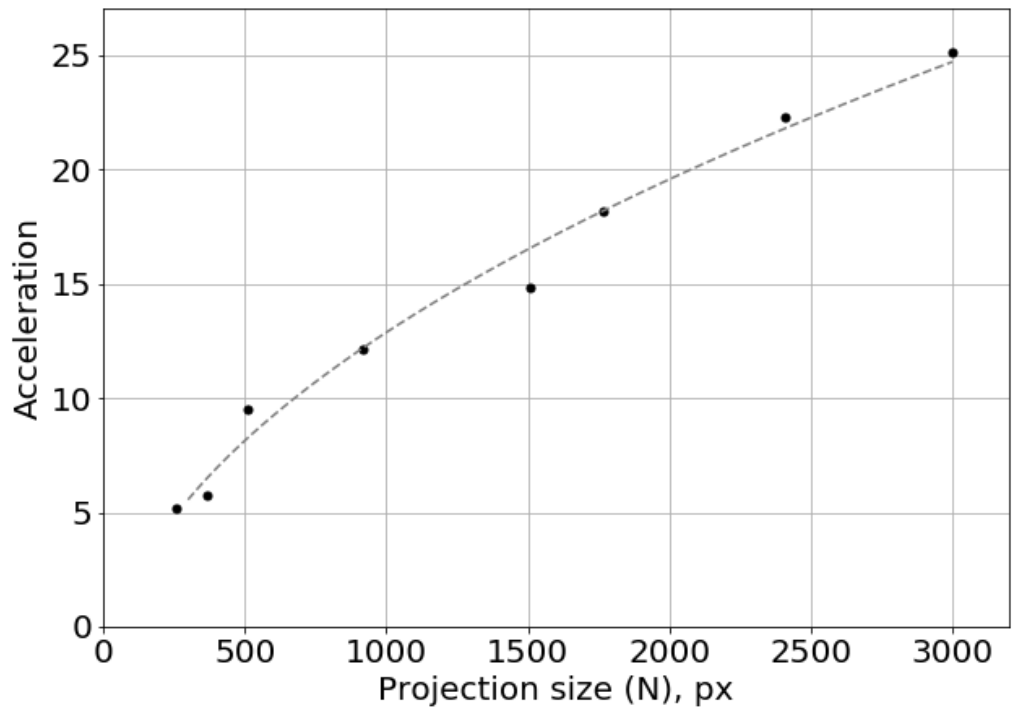}
    \caption{Acceleration of FDK-OD in comparison with FDK as a function of projection width}
    \label{accel}
\end{center}
\end{figure}

\Fref{accel} shows the dependence of acceleration on image size N. Acceleration is defined as the ratio between time consumed by FDK and time that can be obtained with FDK-OD. Dashed line shows that the set of experimental points is roughly consistent with square root function of N. In the measurements of FDK-OD performance one decimation stage was skipped to achieve a similar to FDK image quality.

\section{Quality loss analysis}

It is known that FDK is an approximate algorithm. It can be also expected that FDK with decimations doesn't produce exactly the same image as conventional FDK. Angular decimations and shift interpolations affect the intensity of pixels. It is essential to develop a method that makes it possible to show the difference between algorithms.

\begin{figure}
\begin{center}
  \subfloat[]{
  \begin{minipage}{.49\linewidth}
  \includegraphics[width=\linewidth]{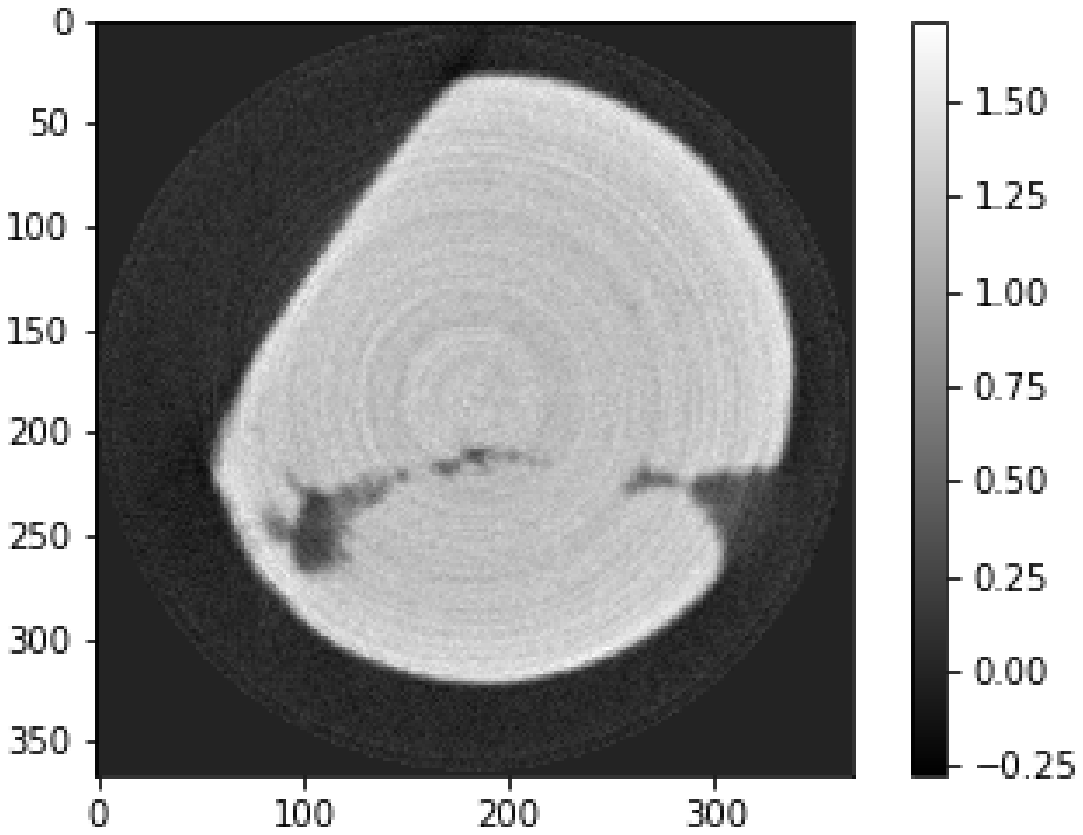}
  \label{comp_fdk}
  \end{minipage}%
  }
  \subfloat[]{
  \begin{minipage}{.49\linewidth}
  \includegraphics[width=\linewidth]{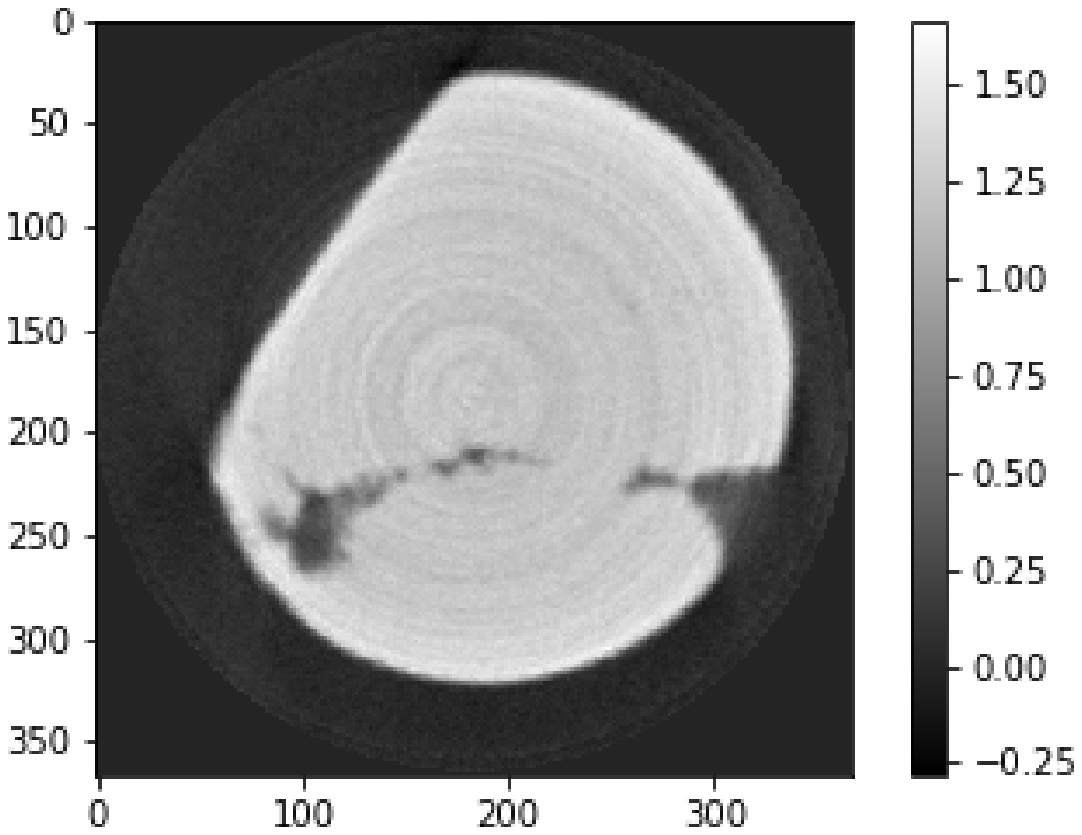}
  \label{comp_fdkod}
  \end{minipage}%
  }\par
  \subfloat[]{
  \begin{minipage}{.49\linewidth}
  \includegraphics[width=\linewidth]{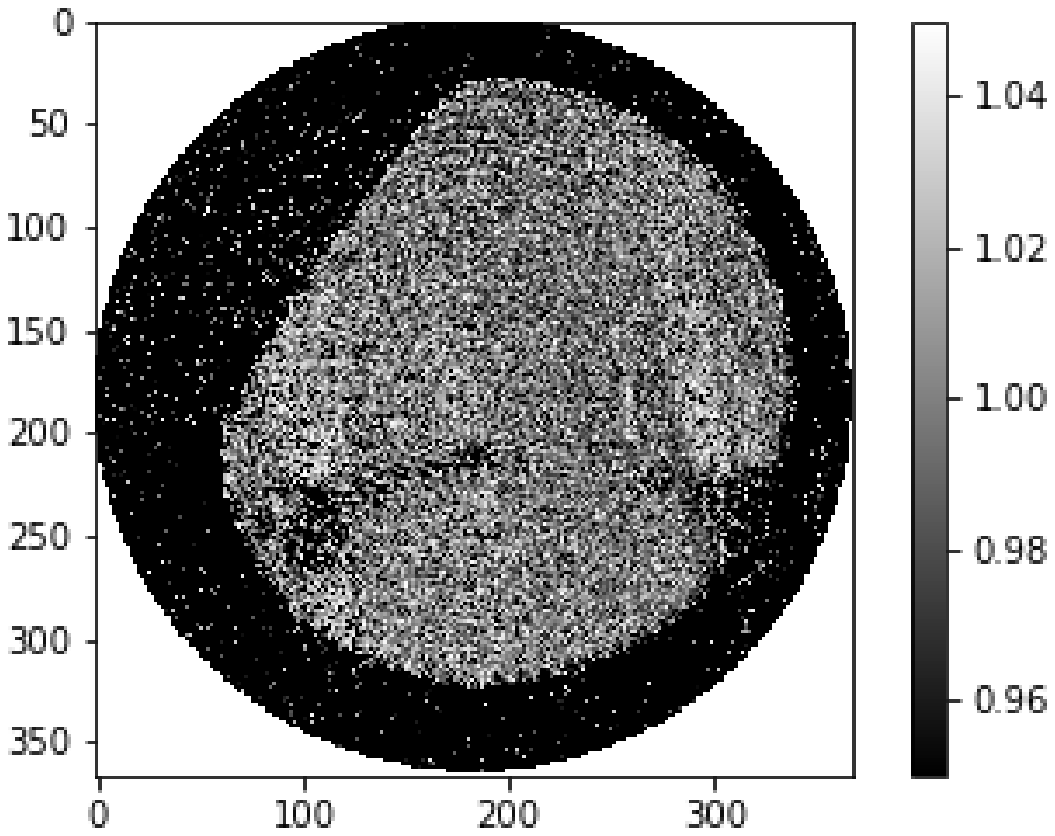}
  \label{comp_div}
  \end{minipage}%
  }
  \caption{Slices reconstructed with the FDK(4a) and FDK-OD(4b), result of pixelwise division(4c)}
\end{center}
\end{figure}

The simplest idea is to divide pixel-wise image, produced by the decomposition algorithm, by image obtained from the FDK. This gives a relative difference between images. An example of using this method is shown in figures \ref{comp_fdk}, \ref{comp_fdkod} and \ref{comp_div}. \Fref{comp_fdk} is a slice reconstructed by the FDK, \fref{comp_fdkod} - by the FDK-OD and \fref{comp_div} shows a result of pixel-wise division of these 2 slices.

In practice, this approach has drawbacks due to several reasons. Angular decimation increases intensity of artifacts on the image background. Intensity of the background pixels is close to zero, so even small divergence leads to big relative difference. Noisy parts of image also present a serious problem. Decomposition stage uses low-pass filter for anti-aliasing and this flattens noise because it is connected to high frequencies. Thus, the simple division mostly shows a degree of similarity for the background artifacts and the noise, that isn't so important for the reconstruction problem.

These considerations support the necessity to found a better way to evaluate the image quality. Practical application require high resolution of image to reconstruct small details of the object. Therefore, it makes sense to analyze the profiles of the border between different parts of the object.

\begin{figure}[h!]
\begin{center}
\subfloat[]{
  \begin{minipage}[b]{0.49\textwidth}
    \includegraphics[width=\textwidth]{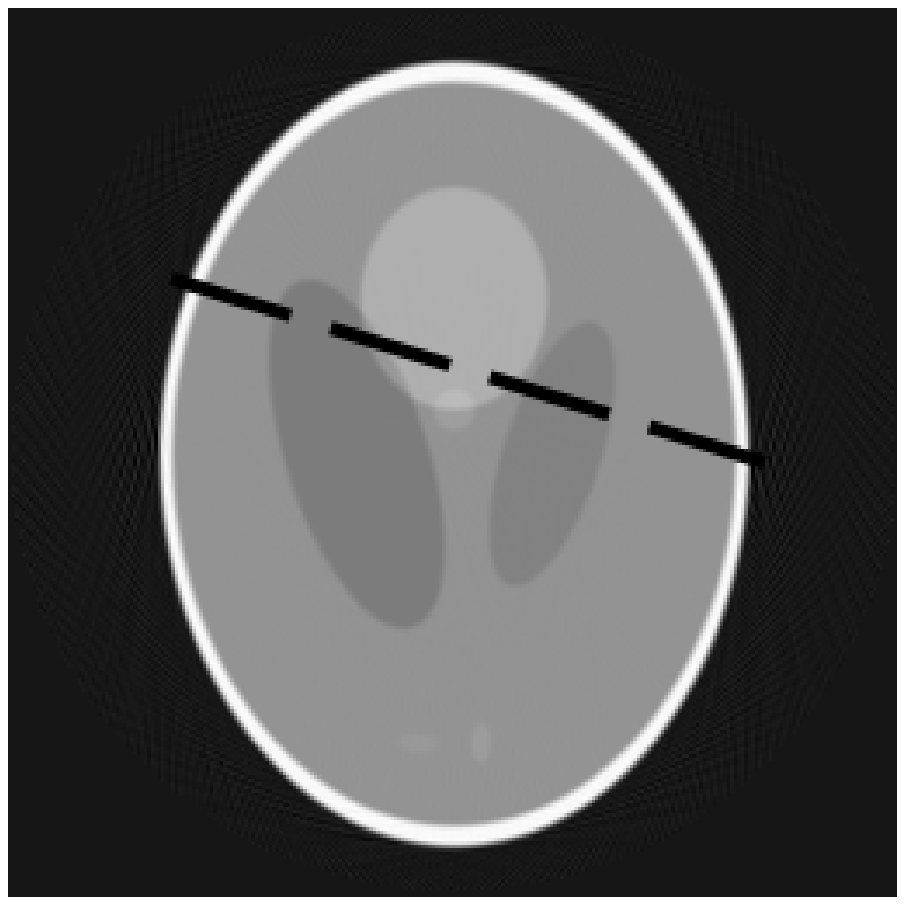}
    \label{shl_fdk}
  \end{minipage}
}
\subfloat[]{
  \begin{minipage}[b]{0.49\textwidth}
    \includegraphics[width=\textwidth]{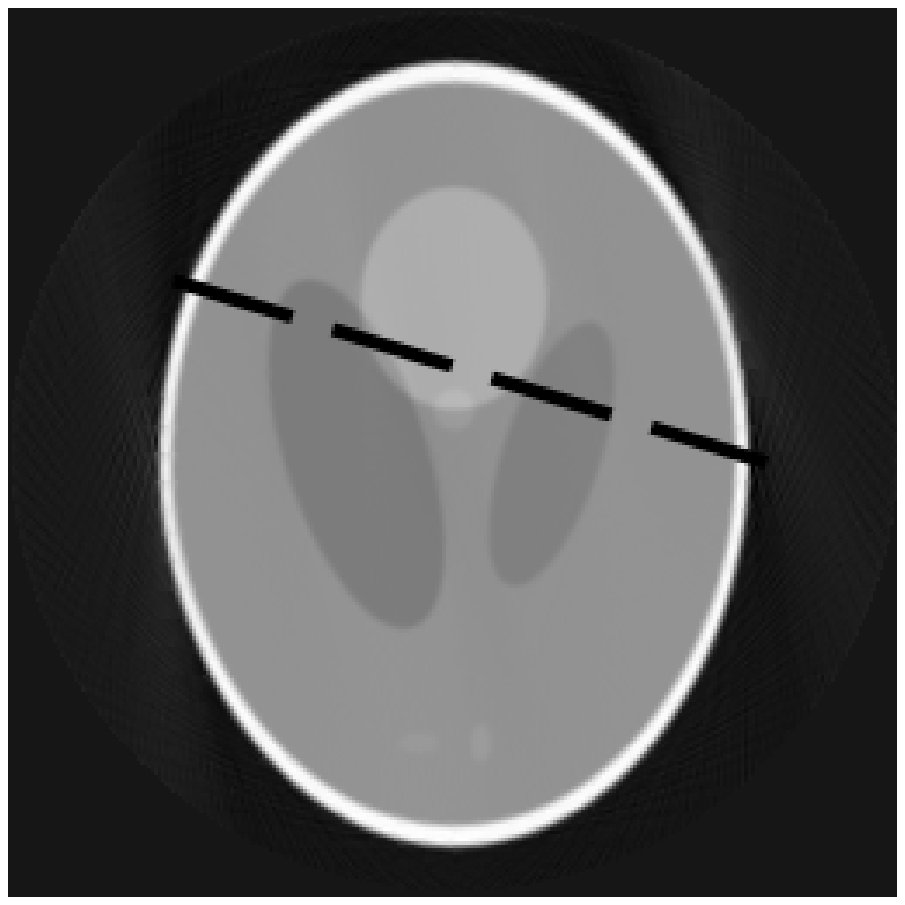}
    \label{shl_fdkod}
  \end{minipage}
}
  \caption{Slices reconstructed by the FDK(5a) and FDK-OD(5b), dashed line corresponds to the profile in the \fref{prof_shl}}
\end{center}
\end{figure}

\begin{figure}[h!]
\begin{center}
	\includegraphics[width=0.9\textwidth]{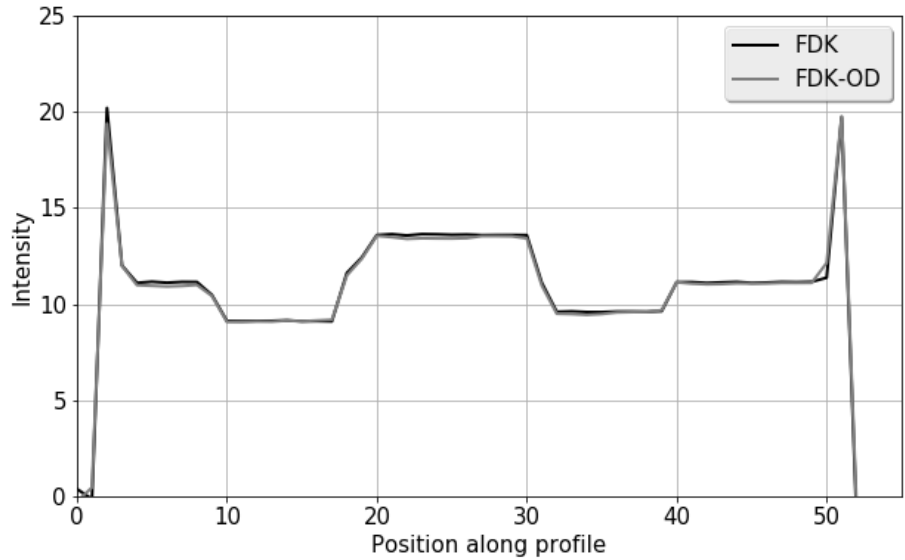}
    \caption{Shepp-Logan phantom profile}
    \label{prof_shl}
\end{center}
\end{figure}

Tests with Shepp-Logan phantom show that decomposition almost doesn't affect uniform regions of slice, thus blurs edges and the borders. Skip of 1 decimation stage significantly decreases the blur and helps to preserve sharp borders. This can be seen on the profile comparison (\fref{prof_shl}) for the objects reconstructed with FDK and with our algorithm (5 stages of decomposition, 1 decimation stage is skipped). Corresponding slices are shown in the figures \ref{shl_fdk} and \ref{shl_fdkod}.

\begin{figure}[h!]
\begin{center}
\subfloat[]{
  \begin{minipage}[b]{0.49\textwidth}
    \includegraphics[width=\textwidth]{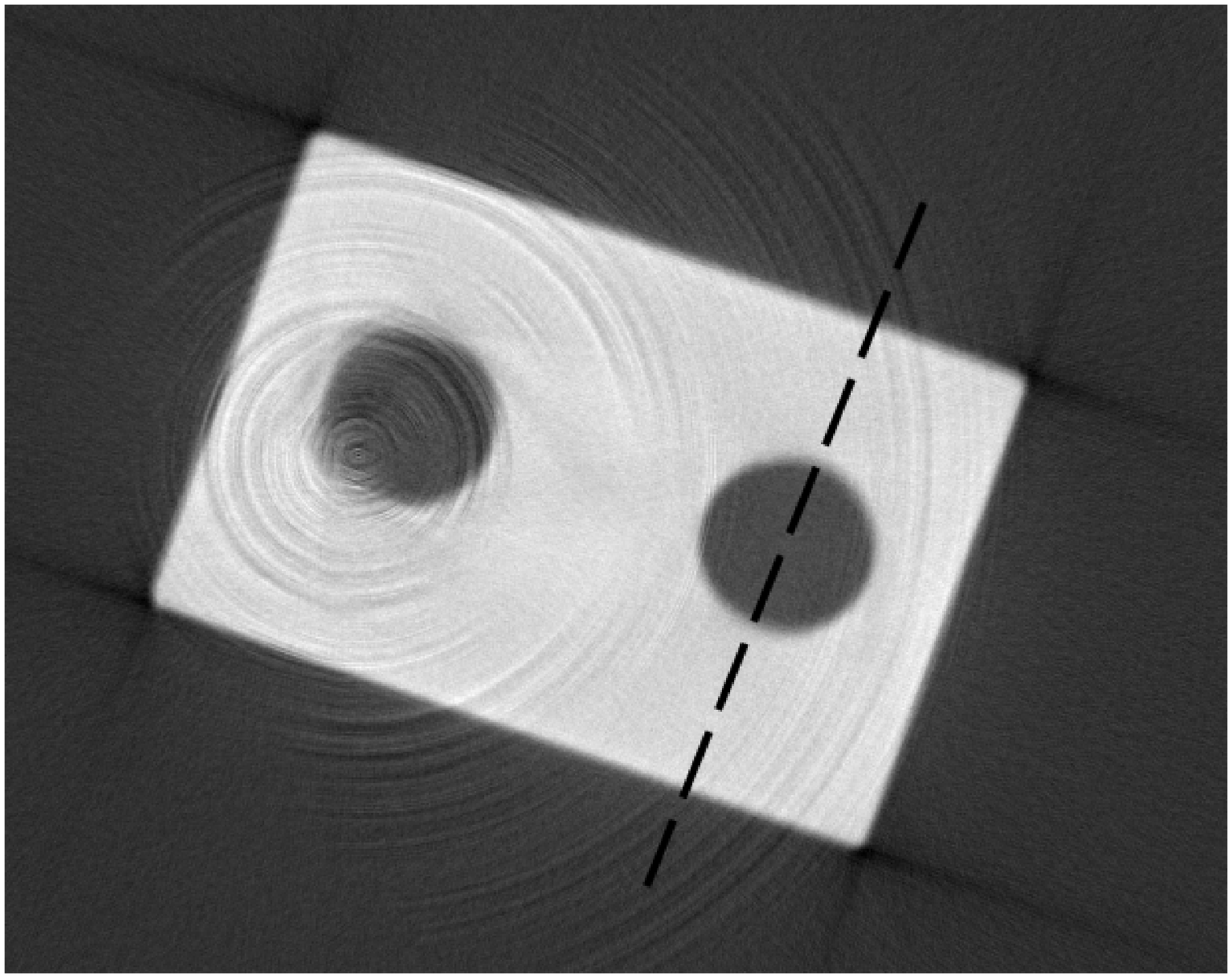}
    \label{real_fdk}
  \end{minipage}
}
\subfloat[]{
  \begin{minipage}[b]{0.49\textwidth}
    \includegraphics[width=\textwidth]{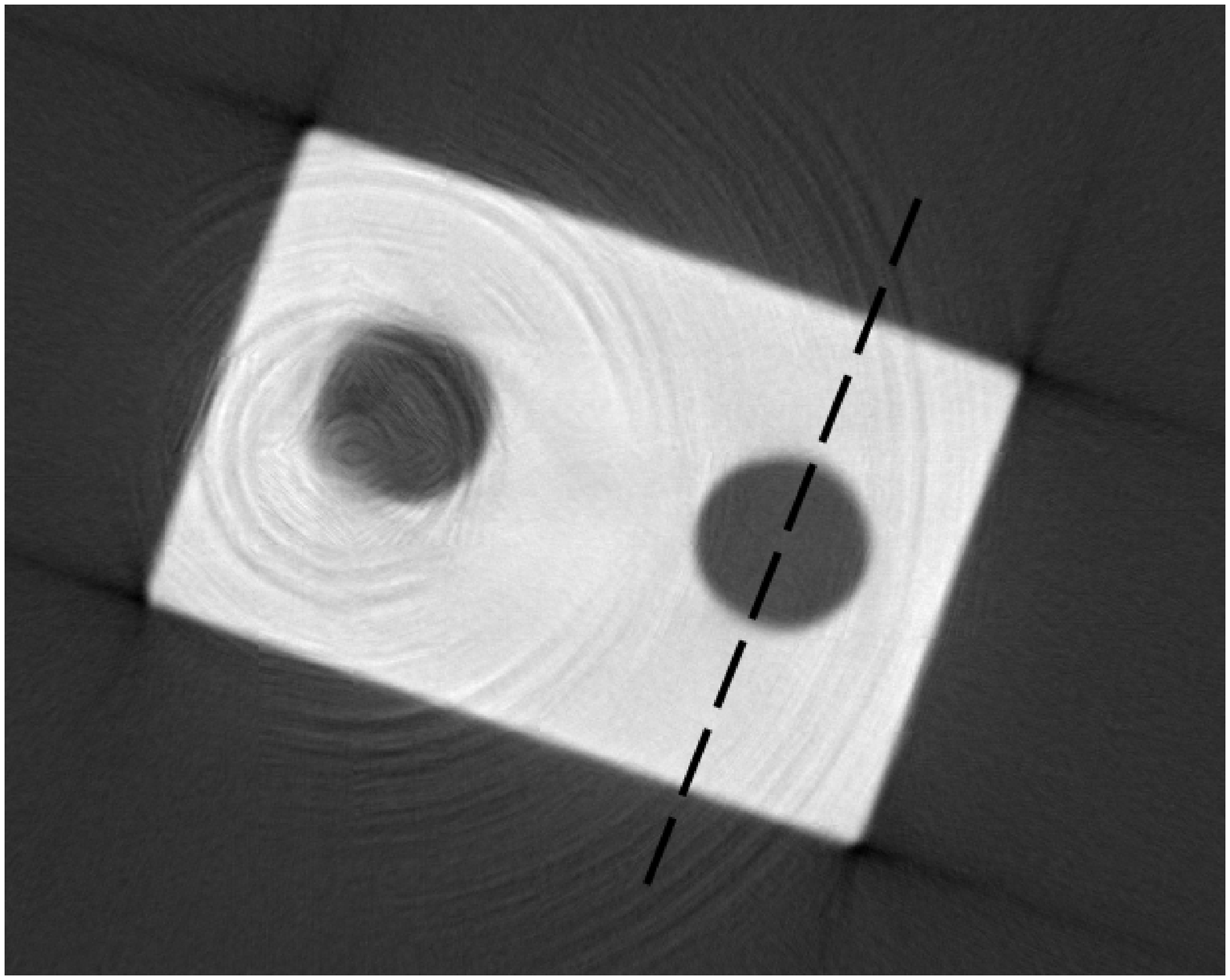}
    \label{real_fdkod}
  \end{minipage}
}
  \caption{Slices reconstructed by the FDK(7a) and FDK-OD(7b), dashed line corresponds to profile below}
\end{center}
\end{figure}

\begin{figure}[h!]
\begin{center}
	\includegraphics[width=0.99\textwidth]{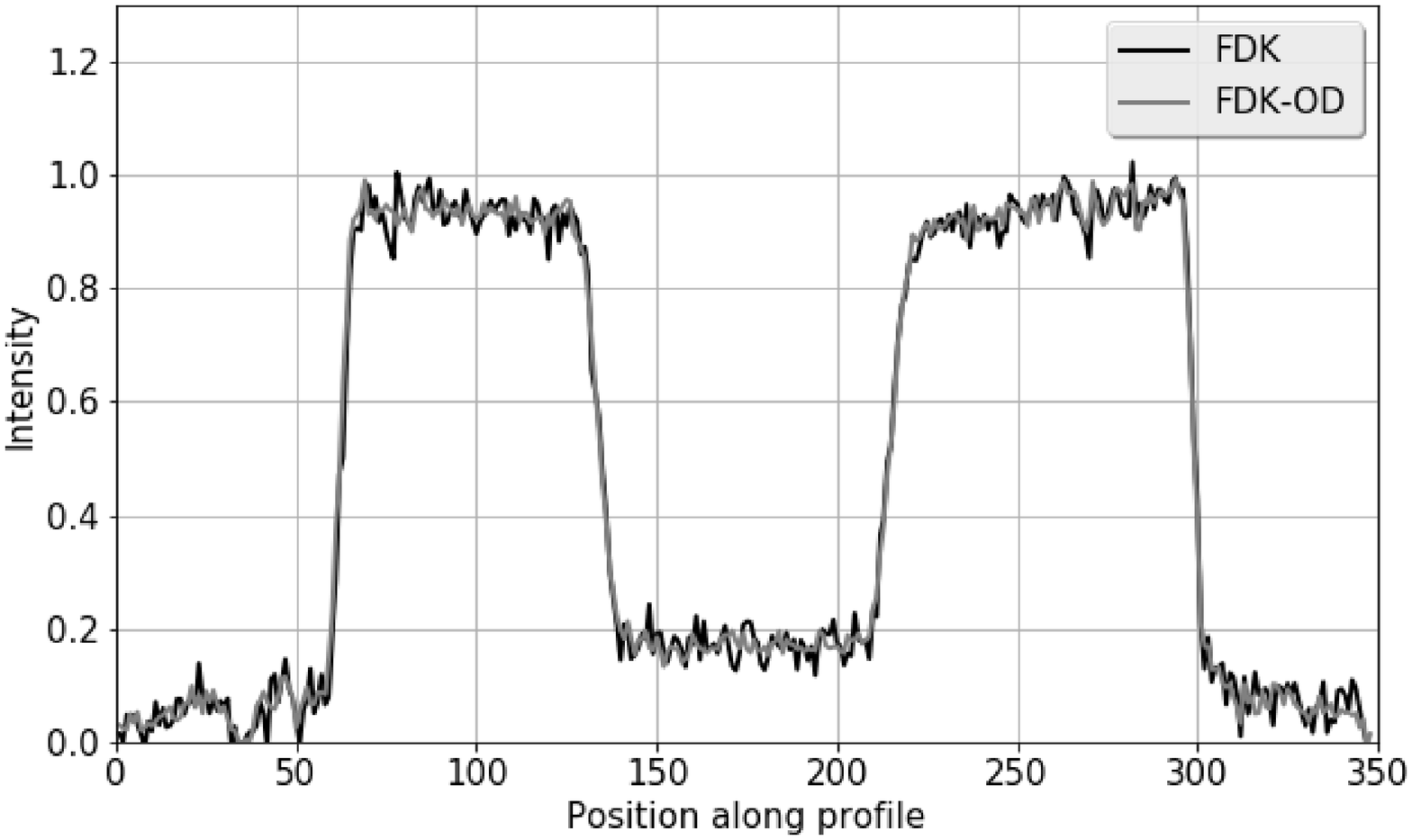}
    \caption{Box object profile}
    \label{real_prof}
\end{center}
\end{figure}

Tests with real data have also been performed. Due to a presence of noise in real data, it can be observed in profiles. Obtained slices are shown in the figures \ref{real_fdk} and \ref{real_fdkod}. The comparison of the profiles (\fref{real_prof}) reveals that the algorithm FDK-OD produces a smoother profile, while preserving sharp borders.

\begin{figure}[h!]
\begin{center}
	\includegraphics[width=0.99\textwidth]{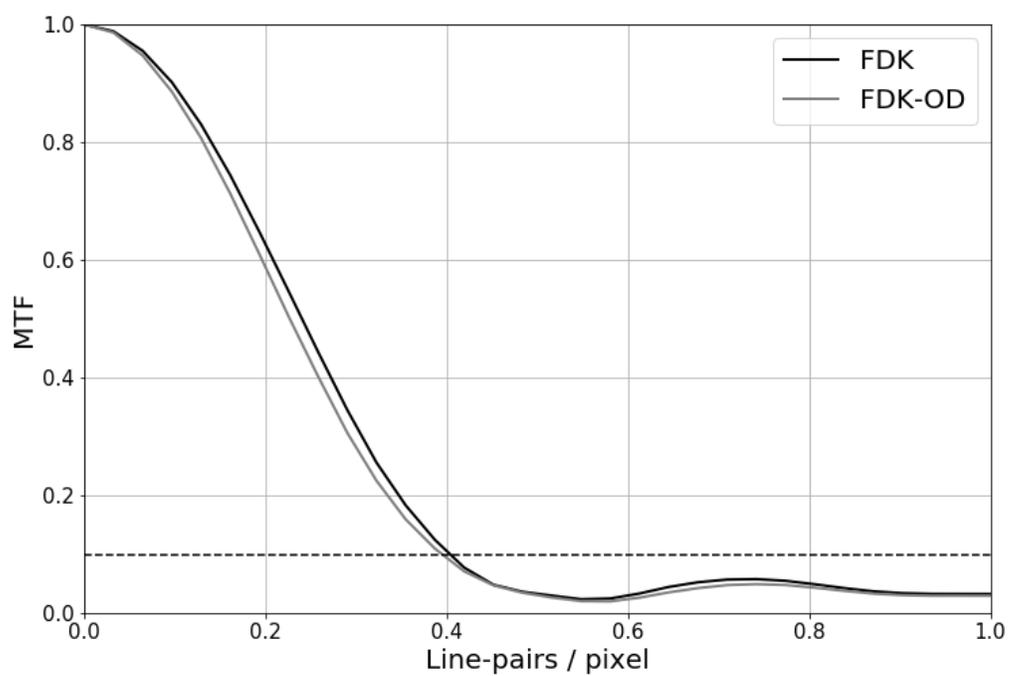}
    \caption{Comparison of MTF for slices reconstructed with FDK and FDK-OD}
    \label{mtf}
\end{center}
\end{figure}

Modulation transfer function (MTF) is widely used to describe the image signal transfer in optical and photographic  systems \cite{mtf}. It can be used to estimate a spatial resolution of reconstructed image. Slanted-edge method of measuring the spatial frequency response \cite{se_mtf} is well known approximation of the MTF and often used in image quality testing. This method analyzes sharp edge and determines how optical system blurs details of object. We applied MTF evaluation to results of simulated box phantom reconstruction after different number of decomposition stages. Figure \ref{mtf} shows that FDK-OD affects the object edges slightly according to the MTF analysis. For ease of comparison, line corresponding to 10\% MTF is specified. 10 \% value is chosen as a border of limiting resolution \cite{dsp}.

\section{Conclusion}

We have presented a new approach to apply decomposition principle to 3D cone-beam reconstructions and demonstrated its suitability. The FDK-OD leads to the computational cost $O(N^{3.5})$ and possesses several interesting features. This implementation gives considerable acceleration without large amount of smoothing operations thereby doesn't require precise interpolation and convolution. Besides, it has potential for significant reduction of peak memory usage. Further studies will be concentrated on computational constant reduction.

\section*{References}
\bibliographystyle{dcu}
\bibliography{main}

\end{document}